\documentclass[doublecol]{epl2} 

\title{Out of Equilibrium Solutions in the $XY$-Hamiltonian Mean Field
model}
\shorttitle{Title} 

\author{X. Leoncini\inst{1} \and T. L. Van Den Berg\inst{1} \and D. Fanelli\inst{2}}
\shortauthor{X. Leoncini \etal}

\institute{                    
  \inst{1} Centre de Physique Th\'eorique\thanks{Unit\'e Mixte de Recherche (UMR 6207) du CNRS, et des universit\'es Aix-Marseille
I, Aix-Marseille II et du Sud Toulon-Var. Laboratoire affili\'e à la
FRUMAM (FR 2291).}, Aix-Marseille Universit\'e, CNRS, Luminy, Case 907, F-13288 Marseille cedex 9, France\\
  \inst{2} Dipartimento di Energetica Sergio Stecco, Universita
di Firenze, via s. Marta 3, 50139 Firenze, Italia e Centro interdipartimentale
per lo Studio delle Dinamiche Complesse (CSDC)
}

\abstract{Out of equilibrium magnetised solutions of the $XY$-Hamiltonian Mean
Field ($XY$-HMF) model are build using an ensemble of integrable uncoupled pendula.
Using these solutions we display an out-of equilibrium phase transition
using a specific reduced set of the magnetised solutions.}

\begin{document}

\maketitle

\maketitle

Long-range interactions are such that the two-body potential decays
at large distances with a power--law exponent which is smaller than
the space dimension. A large number of fundamental physical systems
falls in such a broad category, including for instance gravitational
forces and unscreened Coulomb interactions\cite{Dauxois_book2002}, vortices in two dimensional
fluid mechanics \cite{Onsager49,Edwards74,Weiss91}, wave-particle
systems relevant to plasma physics \cite{ElskensBook2002,Benisti07}, Free-Electron
Lasers (FELs) \cite{Bonifacio90,Barre04,Bachelard07} and even condensed
matter physics for easy-axis anti-ferromagnetic spin chains, where
dipolar effects become dominant. While for short--range systems only
adjacent elements are effectively coupled, long-range forces result
in a global network of inter-particles connections, each element soliciting
every other constitutive unit. Clearly, from such an enhanced degree
of complexity stems the difficulties in addressing the fascinating
realm of long-range systems, for which standard techniques in physics,
notably in the framework of equilibrium statistical mechanics, proves
essentially inadequate. It is in particular customarily accepted that
long--range systems display universal out-of-equilibrium features:
Long-lived intermediate states can in fact emerge, where the system
gets, virtually indefinitely, trapped (the time of escape diverging
with the number of particles), before relaxing towards its deputed
thermodynamic equilibrium. These are the so--called Quasi Stationary
States (QSSs) which have been shown to arise in several different
physical contexts, ranging from laser physics to cosmology, via plasma
applications. A surprising, though general, aspect relates to the
role of initial conditions of which QSS keeps memory. More intriguingly,
such a dependence can eventually materialise in a genuine out-of-equilibrium
phase transition: By properly adjusting dedicated control parameters
which refers to the initial state, one observes the convergence towards
intimately distinct macroscopic regimes (e.g. homogeneous/non homogeneous)
\cite{Antoniazzi2007,Chavanis-RuffoCCT07}. Recently, such phase
transitions for systems embedded in one spatial dimension have been
re-interpreted as a topological change in the single particles orbits
\cite{Bachelard08}. This conclusion is achieved by performing a
stroboscopic analysis of individual trajectories, which are being
sampled at a specific rate imposed by the emerging time evolution
of a collective variable and consequently sensitive to the intrinsic
degree of microscopic self-organisation.

QSSs out of equilibrium regimes have been explained by resorting to
a maximum entropy principle inspired to the Lynden-Bell's seminal
work on the so-called \textit{violent relaxation theory}, an analytical
treatment based on the Vlasov equation and originally developed for
astrophysical applications \cite{LyndenBell67}. 


In this letter we shall 
provide
a strategy to construct a whole family of out-of-equilibrium solutions
with reference to the paradigmatic $XY$-Hamiltonian Mean Field ($XY$-HMF)
model \cite{Antoni95}. This procedure exploits the analogy with
an ensemble made of uncoupled pendula and explicitly accommodate for
self-consistency as a crucial ingredient. Even more importantly, out-of
equilibrium phase transitions are displayed using a reduced set of
the non-homogeneous (magnetised) solutions.
The proposed approach is inspired by the observation that in the continuum limit (for an
infinite number of particles) the discrete set of equations describing
the physical system under scrutiny, converges towards the Vlasov equation,
which governs the evolution of the one-particle distribution function.

The $N$-body Hamiltonian for the $XY$-HMF model with ferromagnetic
interactions writes\begin{equation}
H=\sum_{i=1}^{N}\left[\frac{p_{i}^{2}}{2}+\frac{1}{2N}\sum_{j=1}^{N}1-\cos\left(q_{i}-q_{j}\right)\right]\:,\label{eq:Hamiltonian_HMF}\end{equation}
 where $p_{i}$ and $q_{i}$ are respectively the (canonically conjugate)
momentum and position of particle (rotor) $i$. To monitor the time
evolution of the system, one can introduce the {}``magnetisation''
as \begin{equation}
\mathbf{M}=\frac{1}{N}\left(\sum\cos q_{i},\:\sum\sin q_{i}\right)=M\left(\cos\varphi,\:\sin\varphi\right)\:.\label{eq:Magnetisation_def}\end{equation}
 The equations of motion for the particles can be therefore cast in
the form\begin{equation}
\left\{ \begin{array}{ccc}
\dot{p_{i}} & = & -M\:\sin\left(q_{i}-\varphi\right)\\
\dot{q_{i}} & = & p_{i}\end{array}\right.\:,\label{eq:HMF_motion}\end{equation}
 where the dot denotes the time derivative.
Notice that it is tempting to imagine equations (\ref{eq:HMF_motion}) as resulting
from a set of uncoupled, possibly driven, one dimensional pendula
Hamiltonian. Inspired by this analogy, we here intend to shed light
onto the out-of-equilibrium dynamics of the original $N$-body model,
by investigating the equilibrium properties of an associated pendula
system.
More specifically, let us imagine that the system of coupled rotators
has reached some equilibrium state , such
that in the $N\rightarrow\infty$ limit the magnetisation $M$ of
the $XY$-HMF model is constant and equal to $m$. Equations
of motion (\ref{eq:HMF_motion}) implies that the system formally
reduces to an infinite set of uncoupled pendula. Our strategy to construct
stationary solutions for (\ref{eq:Hamiltonian_HMF}) is to consider a finite ensemble of $N$ uncoupled pendula whose Hamiltonian
reads \begin{equation}
H=\sum_{i=1}^{N}\frac{p_{i}^{2}}{2}+m(1-\cos q_{i})\:,\label{eq:Hamilton_pendulum}\end{equation}
 and compute stationary solutions
in the thermodynamic limit of this $m-$pendula system \cite{Tineke_rapport08}.

\begin{figure}
\includegraphics[width=6cm]{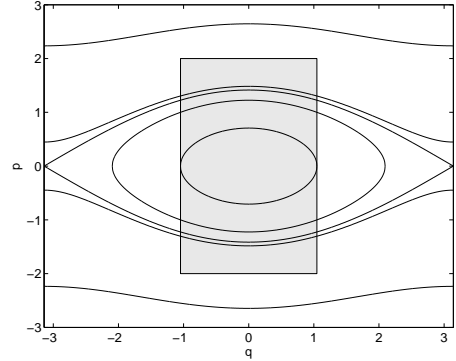}

\caption{Phase portrait of the pendulum for $m=0.5$. The gray region is here used as a tool to estimate numerically the function g(I) as introduced in eq. (9).  It is termed in the literature a 
water-bag and has no reference to the selected particles' initial condition.
\label{fig:Pendulum_waterbag}}

\end{figure}

Given an initial condition each pendulum $i$ is confined on a specific
torus of the pendulum phase portrait depicted in Fig~\ref{fig:Pendulum_waterbag}.%

To build a stationary state we naturally consider the ergodic measure
on the torus which originate from the pendulum motion and time averages.
In order to proceed further in the analysis and due to integrability,
we employ the canonical transformation to the action-angle variables
$(I,\theta)$ of the system (see for instance \cite{ZaslavBook98}).
We thus obtain $H_{m}=H_{m}(I_{i})$, with $\dot{\theta_{i}}=\partial H_{m}/\partial I=\omega(I_{i})$,
where $I_{i}$ stands for the constant action, which is fixed by the
initial state of the uncoupled pendulum $i$. For any selected initial
condition, as time evolves, $\theta$ covers uniformly the circle
$[-\pi,\:\pi[$, while the action $I$ keeps its constant value. The
ergodic measure reduces hence to $\rho_{i}(I,\theta)=\frac{1}{2\pi}\delta(I-I_{i})$,
which immediately translates in \begin{equation}
\rho_{E}=\prod_{i=1}^{N}\rho_{i}\:,\label{eq:Rho_eq_phase_space}\end{equation}
 when considering an ensemble of $N$ particles. Because of the above
and since particles are identical and non interacting one can straightforwardly
deduce the following expression for the one particle density function:
\begin{equation}
f(I,\theta)=\frac{g(I)}{2\pi}\:,\label{eq:One_particle_pdf}\end{equation}
 where $g(\cdot)$ is a discrete valued function, solely determined
by the selected initial conditions. In the limit $N\rightarrow\infty$,
$g(\cdot)$ can change smoothly with the (continuous) action variable.

In order to get a stationary equilibrium solution of the system of pendula 
we can then consider a given positive and integrable function $g$, associate to it the function $f(I,\theta)$
according to (\ref{eq:One_particle_pdf}). Then we perform an {}``inverse''
transform to obtain an explicit expression for the one particle density function $\tilde{f}(p,q)$
as defined in the original phase space $\Gamma_{pq}$.
This latter represents an equilibrium (time invariant)
solution, and  enables us to estimate any macroscopic observable,
defined as a function of the sea of microscopic constituents. More
concretely let us turn to consider the global magnetisation $\mathbf{M}$
as specified by eq. (\ref{eq:Magnetisation_def})). In the infinite
$N$ limit the time average of the magnetisation, which coincides
with the ergodic-spatial average, reads: \begin{equation}
\bar{\mathbf{M}}=\left\langle \mathbf{M}\right\rangle =\left(\int\tilde{f}(p,q)\cos q\: \upd p\upd q\:,0\right)\:,\label{eq:Magnetisation_pendula}\end{equation}
 where the observation that $\tilde{f}$ is even in $p$ and $q$
has been used. Expressing the above in term action-angle variables,
one can write \begin{equation}
\bar{\mathbf{M}}=\left\langle \mathbf{M}\right\rangle =\left(\frac{1}{2\pi}\int g(I)\cos q(I,\theta)\: \upd I\upd \theta\:,0\right)\:.\label{eq:Magnetisation_pendula2}\end{equation}
 The integration over the angle can be performed (see Appendix), yielding
to the final form\begin{eqnarray}
M & = & \int_{0}^{\frac{8}{\pi}\sqrt{m}}g(I)\left(2\frac{E(\kappa)}{K(\kappa)}-1\right)\: \upd I\nonumber \\
 & + & \int_{\frac{8}{\pi}\sqrt{m}}^{\infty}g(I)\left(1+2\kappa^{2}\left(\frac{E(\kappa^{-1})}{K(\kappa^{-1})}-1\right)\right)\: \upd I,\label{eq:Magne_final}\end{eqnarray}
 with $\kappa=\kappa(I)=(H(I)+m)/2m$. Notice that if we consider
an initial distribution given by $\tilde{f}$ , then the magnetisation
(\ref{eq:Magne_final}) stays constant as, by construction $\tilde{f}$
is stationary. In order to reconcile the $m$-model of uncoupled pendula
to the $XY$-HMF interacting rotors, we need to impose the condition
\begin{equation}
\langle\mathbf{M}\rangle=m\:.\label{eq:Implicit_equation}\end{equation}
 Should there exist an $\tilde{f}$ for which equation (\ref{eq:Implicit_equation})
had an $m\ne0$ solution, then we would have obtained a stationary
solution of the system of pendula, which is in turn also magnetised stationary
solution of the $XY$-HMF model in the $N\rightarrow\infty$ limit.
 Indeed the equations of motion for the system of pendula write\begin{equation}
\left\{ \begin{array}{ccc}
\dot{p_{i}} & = & -m\:\sin q_{i}\\
\dot{q_{i}} & = & p_{i}\end{array}\right.\:,\label{eq:Pendulum_motion}\end{equation}
 and given the imposed condition (\ref{eq:Implicit_equation}) they
are formally identical to (\ref{eq:HMF_motion}) with a constant $\varphi=0$.
Note that the condition of the phase  $\varphi=0$, can be modified at will by a simple shift in the 
$m$-pendula Hamiltonian.
\begin{figure}
\includegraphics[scale=0.33]{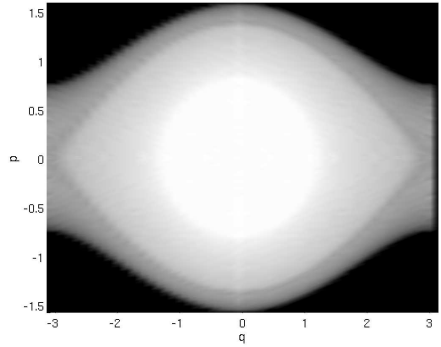}

\caption{Stationary out of equilibrium distribution $\tilde{f}(p,q)$ with
$m=\langle M\rangle=0.5$ obtained for $g(I)$ reconstructed from a water-bag with $q_{0}=2\pi/5$
and $p_{0}=1.37$.\label{fig:stationnary}}

\end{figure}

Such an out of equilibrium solution is displayed in Fig.~\ref{fig:stationnary},
for a specific choice of the initial condition and output magnetisation
amount. One typically recognises the underlying pendulum phase portrait,
with each tori being differently populated according to the function
$g(I)$. The non-uniformity of $\tilde{f}(p,q)$ on each torus stems
from the nonlinearity of the transformation $q=q(I,\theta)$, $p=p(I,\theta)$.

As previously mentioned out of equilibrium, phase transition have
been previously reported separating macroscopically distinct QSSs
phases. These findings are here revisited in the framework of the
proposed approach, which holds promise to generalise the conclusion
beyond the water-bag regime so far inspected via the Lynden-Bell ansatz.
The water-bag regimes  correspond  to initially assign the particles
to populate, randomly and uniformly, a bound domain $[-q_{0},\: q_{0}]\times[-p_{0},\: p_{0}]$
depicted in gray in Fig.~\ref{fig:stationnary}.
This latter initial condition is then univoquely specified by the magnetisation at time zero, namely $M_{0}=\sin(q_{0})/q_{0}$, and the energy per particle $U=p_{0}^{2}/6+(1-M_{0}^{2})/2$.
Notice that the water bag concept will be here invoked as a mere numerical strategy to calculate the needed 
function $g(I)$. We are hence not limiting the present analysis to a specific  class of initial conditions, as e.g. done in \cite{Antoniazzi2007}, but we rather present a compelling evidence on the existence of a phase transition in a broader perspective.

Equation (\ref{eq:Implicit_equation}) is implicit in $m$ parametrised
through the initial conditions which enter the definition of the function
$g$. Such an equation admits $m=0$ as a trivial solution. One can
then look for more general solutions of Eq.~(\ref{eq:Implicit_equation})
with $m\ne0$. Even though $\tilde{f}(p,q)$ water bag type is definitely not of, the water-bag type, as clearly depicted in Fig.~\ref{fig:stationnary}, we have here decided to facilitate the forthcoming analysis, namely the calculation of the associated  $g(I)$, by focusing on a finite portion of phase space as delimited by a water-bag window.
 This choice 
allows us to obtain
a family of solutions monitored by the two same parameters, namely $U$ and $M_0$.
To  construct $g(I)$ it is possible to analytically compute the length of the intersection of each tori with the water-bag (see Fig.~\ref{fig:stationnary}), however we settled for a simple numerical procedure.
Namely we consider a large ensemble of particles whose distribution approximates
a waterbag, then for each particle and a given $m$ we compute its
corresponding action to construct an histogram of $I$ and use it
as an approximate form of $g(I)$. Finally we use this expression
and check whether $m=\langle M\rangle$ and look for possible solutions.
The drawback of this choice results 
in a dependence
on the number $N$ of particles used to construct
the approximate waterbag wich may be important when we are close to the $m=\langle M\rangle=0$
transition line. We however believe its accuracy is sufficient to present our point.%
\begin{figure}
\includegraphics[width=7cm,keepaspectratio]{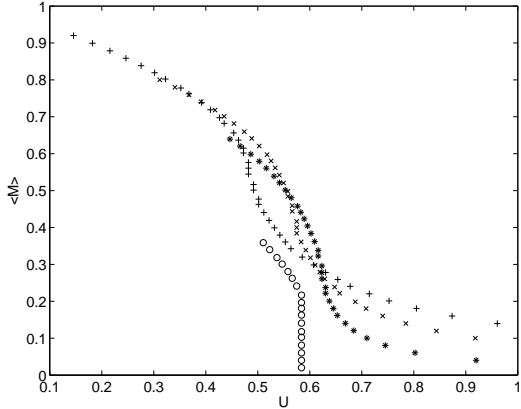}

\caption{Magnetisation $\langle M\rangle$ as a function of energy density
$U$ for different values of $M_{0}$. The symbols $+,\:\times,\:+,\:\circ$
correspond to non-zero solutions of the system of equations  (\ref{eq:Magne_final}) and (\ref{eq:Implicit_equation}) computed respectively for $M_{0}=0.854,\:0.637,\:0.368,\:0$. One
can notice a dependence on $M_{0}$, as well as a first order transition
for $M_{0}=0$. Computation of the function $g(I)$ is made using
an approximate water-bag filled with $1.6\:10^{7}$ particles. Note also that the that the transition point around $U=0.59$ for $M_0=0$ is very close to that found, numerically in \cite{Antoniazzi2007} probably because the initial condition is  ``closer'' to a water-bag one when $M_0=0$.  \label{fig:M-vs-U}}

\end{figure}

Results for different $M_{0}$ are depicted in Fig.~\ref{fig:M-vs-U}.
One can clearly appreciate the transition from a magnetised state
to a non magnetised one, as well as a first order type of transition
for $M_{0}=0$.

A few comments are mandatory at this point. First, we insist on the
fact that the phase space trajectories corresponding to derived solutions
and its associated time evolution in $\Gamma_{pq}$ apply to two,
intrinsically different, dynamical systems, namely (\ref{eq:Hamiltonian_HMF})
and (\ref{eq:Hamilton_pendulum}). Second, consider the average energy
per particle $U=E/N$. We point out that the constants in Hamiltonians
(\ref{eq:Hamiltonian_HMF}) and (\ref{eq:Hamilton_pendulum}) are
chosen in order to have $0$ as a minimal value for the energy, which,
in the thermodynamic limit, implies $U=0$ for a zero temperature.
Now focus on $U_{HMF}$ for the $XY$-HMF model: We obtain $U_{HMF}=T/2-M^{2}/2+1/2$,
while for the pendula one gets $U_{p}=T/2-M^{2}+M$. Here, in both
cases, $T/2$ is the average kinetic energy per particle. Different
energies are thus associated to the same trajectory, depending on
the dynamical system that is being chosen to generate it. In order
to reconcile the two models one can infer that the chemical potential
of the particle is different, yielding to $\delta\mu=\delta U=(M-1)^{2}/2$
for respectively the integrable uncoupled model and the globally coupled
one. 
Moreover, solutions with constant $M=m=0$ do correspond
to a one dimensional perfect gas: the observed phase transition can
hence be seen as a sort of sublimation.

In conclusion, with reference to the $XY$-HMF model, we have designed
an analytical scheme which allows to identify all possible stationary
solutions with constant magnetisation $\mathbf{M}$ using a set of integrable uncoupled pendula.
This includes as a subset the celebrated QSSs, which are therefore formally understood
within a consistent mathematical framework. Following these lines,
it can be inferred that the out of equilibrium states predicted by
the statistical mechanics scenario pioneered by Lynden-Bell \cite{Chavanis-RuffoCCT07},
should belong to the class of solutions here displayed. We are then
providing de facto a testbed for accuracy of the controversial Lynden-Bell
theory \cite{LyndenBell67}. Note though that the stability of the solutions has
not been discussed, it is currently under investigation and is likely to provide further restrictions on the possible out of equilibrium stationary states.

\acknowledgments
We acknowledge usefull conversations with B. Fernandez, S. Ruffo and M. Vittot. X. Leoncini thanks the Center for Nonlinear Science at Georgia Tech for their hospiltality and support during part of this work and acknowledges
 partial support from the PICS program of the CNRS number 4973.

\section{Appendix}

We shall here review the main mathematical tools which are employed
in the above derivation. When it comes to the pendulum motion, trapped
orbits (libration) are characterised by: \begin{eqnarray*}
q & = & 2\sin^{-1}\left[\kappa sn\left(\frac{2K(\kappa)\theta}{\pi},\kappa\right)\right]\\
I & = & \frac{8}{\pi}\sqrt{m}\left[E(\kappa)-\kappa'^{2}K(\kappa)\right]\\
\langle\cos q\rangle & = & 2\frac{E(\kappa)}{K(\kappa)}-1\end{eqnarray*}
 while for the untrapped ones (rotation) the following relations hold:
\begin{eqnarray*}
q & = & 2am\left(\frac{2K(1/\kappa)\theta}{\pi},\kappa^{-1}\right)\\
I & = & \frac{8\kappa}{\pi}\sqrt{m}E(1/\kappa)\\
\langle\cos q\rangle & = & 1+2\kappa^{2}\left(\frac{E(\kappa^{-1})}{K(\kappa^{-1})}-1\right)\end{eqnarray*}
 with $\kappa^{2}=(h+m)/2m$. Here $<\cdot>=\int_{0}^{2\pi}\cdot \upd\theta/(2\pi)$.
\bibliographystyle{eplbib}

\end{document}